\pdfoutput=1
\pdfoutput=1
\pdfoutput=1
\pdfoutput=1
\pdfoutput=1

\documentclass[fleqn
]
{article}

\usepackage[T2A]{fontenc} 
\usepackage{amsfonts,amssymb,amsmath, latexsym, mathtext} 
\usepackage{bm}
\usepackage[all]{xy}
\usepackage{textcomp}
\usepackage{graphics,graphicx,epsfig}
\usepackage{color}
\usepackage{draftwatermark} 
 \SetWatermarkScale{2.3}
 \SetWatermarkLightness{1}
 \SetWatermarkText{\textbf{Watermarked}}

\usepackage[cp1251]{inputenc}
\usepackage [ english
] { babel }
\usepackage{makeidx}
\makeindex

\usepackage{xr-hyper}
\usepackage[unicode,pagebackref]{hyperref}

\newcommand{\beq}{\begin{equation}}
\newcommand{\eeq}{\end{equation}}
\newcommand{\beqa}{\begin{eqnarray}}
\newcommand{\eeqa}{\end{eqnarray}}

 \title{ 
On the steganographic  image based approach 
to PDF files
 protection
 }
      \author{V.N. \ Gorbachev,   L.A.\ Denisov, E.M.\ Kaynarova \footnote{helenkainarova@gmail.com},   I.K.\ Metelev.
      \\
      \\
      \textit{High School of Printing and Mediatechnoilogies} \\
     \textit{ St-Petersburg State University of Industrial Technology and Design}} 
\makeindex
\begin{document}
\maketitle
\begin{abstract}
Digital images can be  copied without authorization and have to be protected. Two schemes for watermarking images in PDF document were considered. Both schemes include a converter to extract images from  PDF pages and return the protected images  back.  Frequency and spatial domain embedding were  used for hiding a message presented by a binary pattern. 
We considered visible and invisible watermarking and found that spatial domain LSB technique can be  more preferable than frequency embedding 
using  DWT. 
\end{abstract}

Key words: image protection, watermarking, PDF, wavelet,  least significant bit.

\section{Introduction}

The wide spread PDF  has  the powerful  cryptographic tools  to protect information in PDF documents. However there is no unique perfect method of protection and  various  solutions  are proposed in this field.
\\
The rich structure of PDF allows us to use steganographic techniques 
for embedding visible and invisible marks that  protect digital data.
Numerous approaches consider varying  lines or words,  spacing, font characteristics as well as   varying certain invisible characters  as a cover to hide a secrete message.
The developed  White Space Coding technique is based on the fact  
that there are a lot of white-space characters  separating syntactic constructions  one from  another \cite{Tsai}. Visually they are undistinguished or invisible and may be used as covers.  As a result a message bit can be encoded by  two items:  as the normal or as a  non-breaking space. White Space Coding may include a version of RSA encryption on quadratic residual \cite{3wmpdf}.  This modification is suitable for secrete communication via PDF files when a message is encoded by a between-word character location.
Additionally such operators   as  the justified text operator TJ 
can  be considered for data hiding  \cite{TJ}. 
Indeed, we often make the text  justified so that the right margin is not ragged. This edit format results in 
random  position of each character.  A stream of such positions generated by a TJ operator  is a good cover for 
data
hiding
 using  LSB (Least Significant Bit) embedding  \cite{Repo}.
\\
An example is a 
(t,n) secrete sharing scheme. It may be used to share  secrete images  
\cite{SSimage} and to protect  PDF files \cite{SSpdf}.
To protect a very important PDF file it is decomposed into $n$ parts. Each of them is embedded into its PDF cover. All stego covers look equally.         
So, the PDF file is shared among $n$ parties and any $t$ parties  can recover it  if they will cooperate.  
Indeed,  a standard option such Attach File, available at Adobe Acrobat Pro can be suitable
for embedding  \cite{SSpdf}.
\\  
A digital  watermark can be encoded by a self inverting permutation, i.e. a permutation which cycles have  length less or equal 2.  
Using a particular representation,
named 1d and 2d, 
the permutation is embedded into a 1d and 2d array. This technique was developed for audio files and images and generalized for PDF files \cite{Permutpdf}. 
The key idea is that the self-inverting permutation allows to locate the watermarked area to recover the information instead of to extract the message directly.
In the case of frequency domain watermarking this method is robust to JPEG lossy compression \cite{PermutwmFreq}.  
\\
Numerous types of the PDF files raise  a large number of solutions. 
 \cite{ReferySuggest}.
Indeed, now the Adobe Acrobat has options to embed a visible watermark. 
It protects  important documents sometimes printed on a background containing large gray digits or such  inscription as ''watermarked''. This makes the content perception worse and the interested user can remove the watermark using a key for an additional fee.  However Adobe Acrobat uses a simple technique and  numerous suggestions on how to delete the embedded data can be found in the Internet. 
\\
\\
In our paper we focus on protection of images placed in  a PDF files and considering two steganographic schemes. The key idea is to extract 
the desired image and return it back in the file after embedding watermark. 
We introduce a convertor that transforms PDF to SVG format and back \footnote{Available at 
http://pdf.welovehtml.ru}. 
SVG  stores the image in PNG format, so 
a png cover is prepared. The aim of our paper is to investigate  
a spatial and frequency domaine watermarking for images in PDF file. 
We consider LSB and DWT techniques for invisible and visible and removal watermarks.
It was  found that the embedding in a spatial domain is more preferable. 
It introduces less degradation in  case of visible and invisible 
watermarking. 
For spatial watermarking we found a nice quality of the image retrieved after removing visible watermark, its degradation 
is less than for similar DWT technique developed in 
\cite{LessPSNR}.
\\
The paper is organized as follows.  First we consider our scheme including convertor, then introduce the frequency and spatial domain embedding schemes.

\section{Scheme}
\begin{figure}[!h]
  \begin{center}
    \includegraphics[width=13cm,height=18cm]{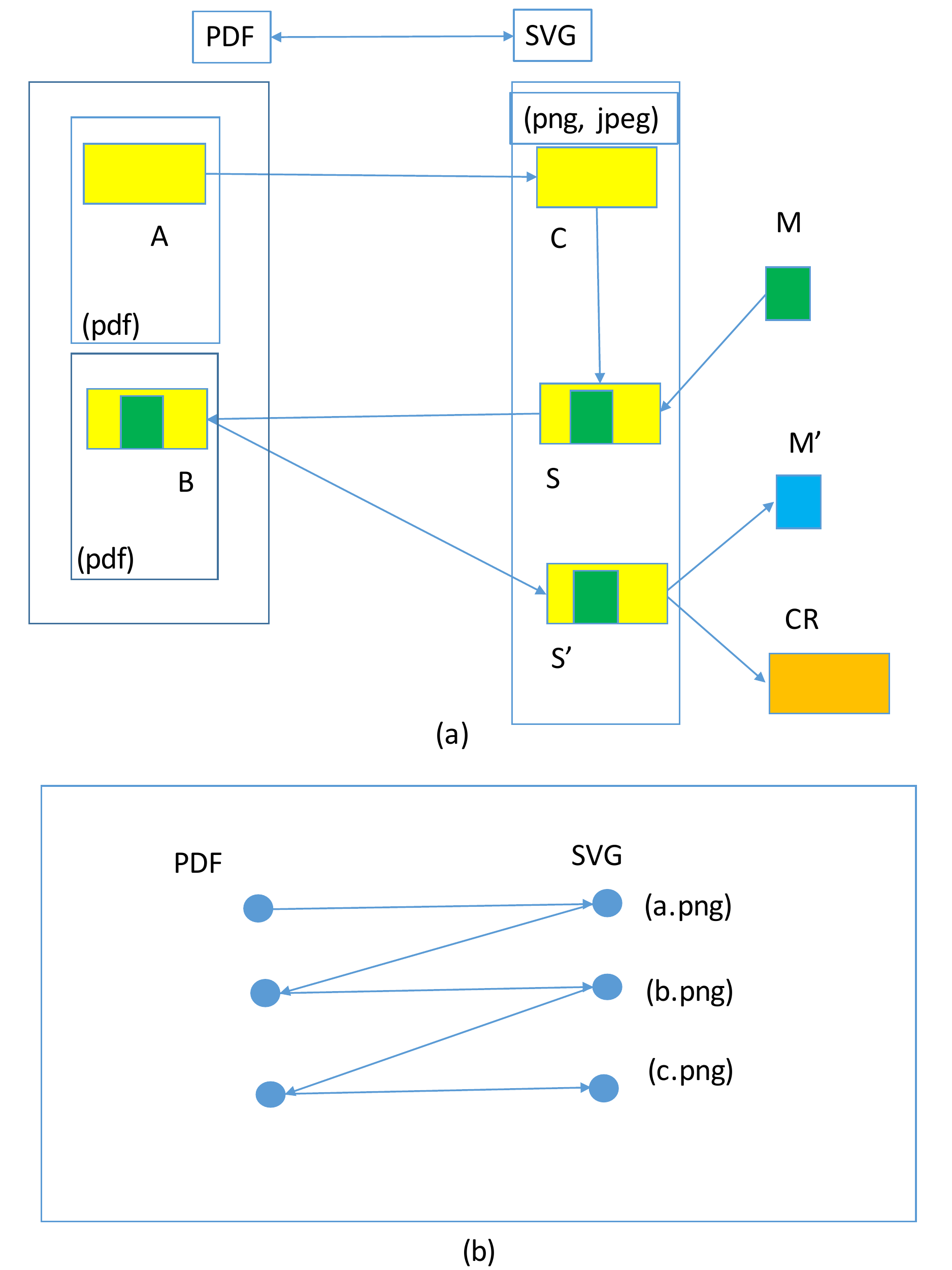}
 \end{center}
 \caption
 { Image-based watermarking of PDF.  a) Image $A$ is extracted from PDF document by convertor $PDF-SVG$, the image is stored in PNG format and watermarked. Convertor returns  the watermarked image back. Two  pattens  $M$ and $M'$ are embedded and extracted watermarks, $CR$ is the retrieved original after removing watermark. 
b)  Conversion  $PDF-SVG$. 
}
 \label{u1}
 \hyperlink{u1}{W}
\end{figure}

\textbf{Setup.}
The scheme has a PDF-SVG convertor and watermarking algorithms, that  allows us to embed invisible  watermarks into image of PDF document  and also visible watermarks that can be removed.  It works as follows.
\begin{itemize}
\item The convertor extracts an image from the PDF document
\item a watermark is embedded (detected/ removed)  into the extracted image
\item the convertor returns  the image back into the PDF document. 
\end{itemize}  
\hypertarget{u1} 
Some details are presented in  Figure (\ref{u1}) (a). 
 An image  $A$ is extracted  from PDF and stored in PNG or JPEG format using  the PDF-SWG convertor. 
So we get a  cover image  $C$ that is watermarked.   We consider a binary image $M$ as a watermark The watermarked image  $S$ is returned back  into PDF by the convertor. The protected image $B$  appears in the PDF document. 
To get the embedded information the convertor extracts $B$,  stores it in graphical format and either detect watermark $M'$ or retrieves cover work as $CR$. The  obtained  $M'$ and $CR$ may differ from its originals because of errors caused  by numerous transformations. So,  the  problem  
is to achieve the indistinguishability
\begin{eqnarray*}
\label{01}
&& M\approx M',\\
&&CR\approx C. 
 \end{eqnarray*} 
\textbf{Distortion measures.}   
To analyze the scheme we introduce the distortion measures 
between the original and the extracted watermarks $d(M,M')$  and  between the original 
and the retrieved cover images $d(C,CR)$. We will use  standard measures as 
RMSE (Root Square Error), PSNR (Peak Signal Noise Ratio) and relative entropy known also as Kullback - Leubler entropy.  For particular case of two binary images $M$ and $M'$ we introduce Hamming distance $ham(M,M')$, There are two reasons for it. The first one says that  PSNR and RMSE are not well agreed with visual perception for binary images. The second one says that, Hamming distance has a clear meaning, that is number of errors or number of different bits in two messages.    
\\
So we introduce the Hamming distance 
\begin{eqnarray}
\label{02}
&&ham(M,M')=(1/\Omega)\sum_{mn}h_{mn},
\end{eqnarray}
where the matrix of errors $h_{mn}$ is defined by
\begin{eqnarray*}
&&h_{mn}=1, \ \  M_{mn}\neq M'_{mn}.\\
&&h_{mn}=0, \ \ M_{mn}= M'_{mn}.
\end{eqnarray*}
where $\Omega$ is a total number of pixels. It follows from 
(\ref{02}) that $ham(M,M')$ is the relative number of errors in the extracted watermark $M'$.  
In  case of binary date Hamming distance can be expressed by 
PSNR or Euclidian metric 
\begin{eqnarray*}
&&RMSE(M,M')=  \sqrt{(1/\Omega)\sum_{mn}(M_{mn}-M'_{mn})^{2}}
\end{eqnarray*}
If $M$ and $M'$ are binary matrix, then
\begin{eqnarray*}
&&RMSE^{2}(M,M')= ham(M,M').
\end{eqnarray*}
\textbf{Convertor.} 
The proposed convertor transforms  PDF document 
into a set of SVG, PNG, JPEG files and back.
This solution allows us to introduce the steganographic  techniques for image watermarking. We need a reversible convertor not to loose information. 
Let us consider the transformation presented in  Figure
(\ref{u1}) (b)
\begin{eqnarray*}
&& \mathsf{PDF\to a.png\to PDF\to b.png\to PDF\to c.png},
\end{eqnarray*}
 where image $a.png$ is extracted from the PDF document and  converted step-by-step. Operations  are reversible if
\begin{eqnarray}
\label{02}
&& \mathsf{ a.png= b.png=c.png}.
\end{eqnarray}
Indeed, in the Internet there are a large number of services for converting  PDF to PNG and  PNG to PDF. Two random convertors were chosen 
\footnote{http://pdf-png-jpg.eu/, http://online2pdf.com/convertor-png-to-pdf 
} 
 and a PNG image was processed. It was a standard 8-bit grayscale image in integer coding  $u_{8}=0,1,\dots, 255$.  The difference 
$a-b$ was found to be up to 129. Such difference in the pixel brightness  can be visible in the  eye.
\\
We checked (\ref{02}) for our convertors using 20  pdf documents and found  all png images
$\mathsf{a,png}$, $\mathsf{b.png}$ and $\mathsf{c.png}$ to be equal.   
It tells us that our convertor is reversible at least for  date in integer encoding. 
This is important for our scheme because errors should be introduced by embedding algorithms and any transformations except PDF-SVG conversion.  
 
\section{Frequency DWT embedding}

\begin{figure}[!h]
  \begin{center}
    \includegraphics[width=13cm,height=18cm]{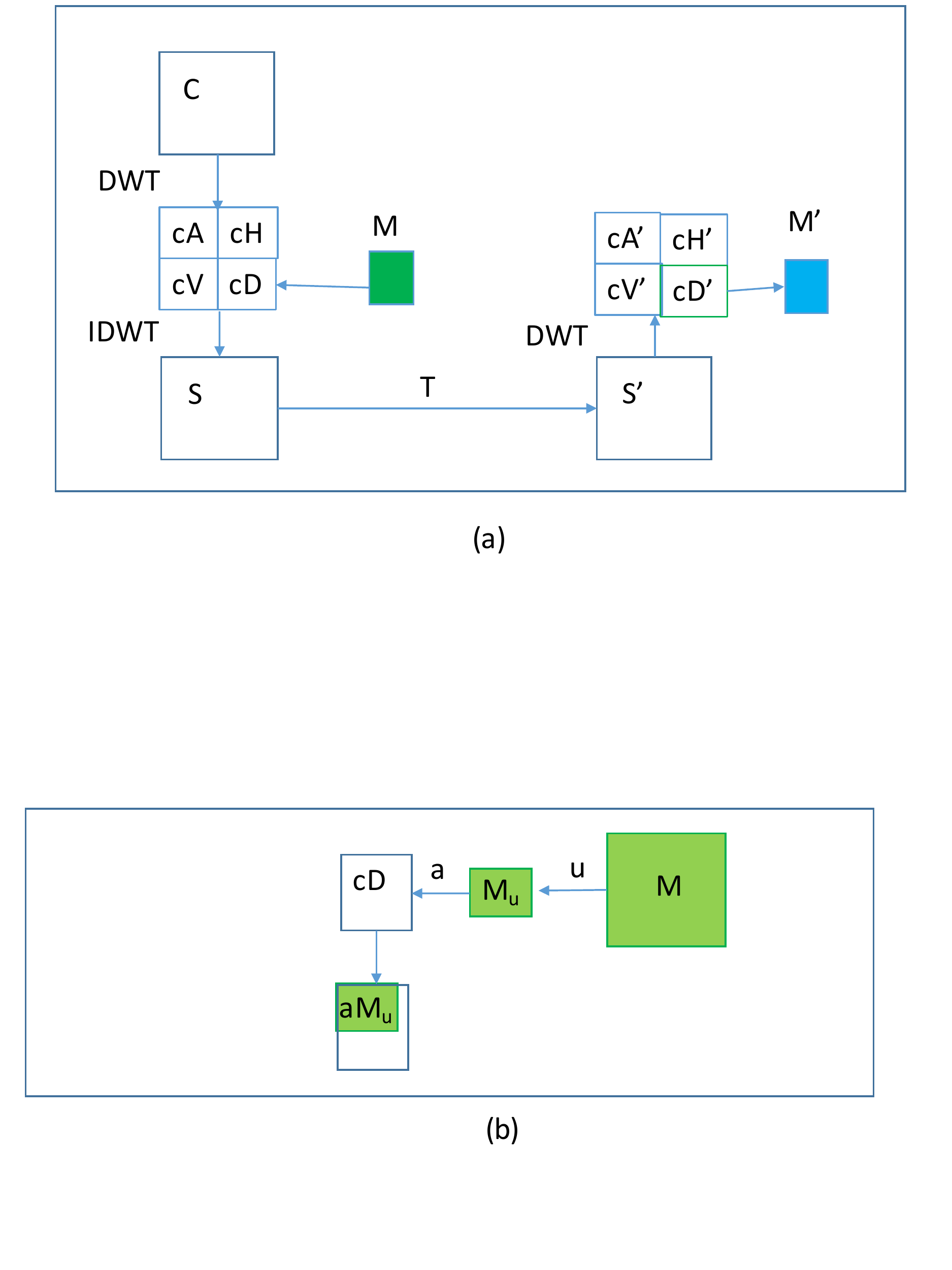}
 \end{center}
 \caption
 { Frequency domain embedding.  
 a)  DWT coefficients of $cD$ block are replaced which  binary image $M$, a watermark, that is detected after some transformations $T$.  
b)  Scaling of watermark  
}
 \label{u2}
 \hyperlink{u2}{W}
\end{figure}

\begin{figure}[!h]
  \begin{center}
    \includegraphics[width=13cm,height=18cm]{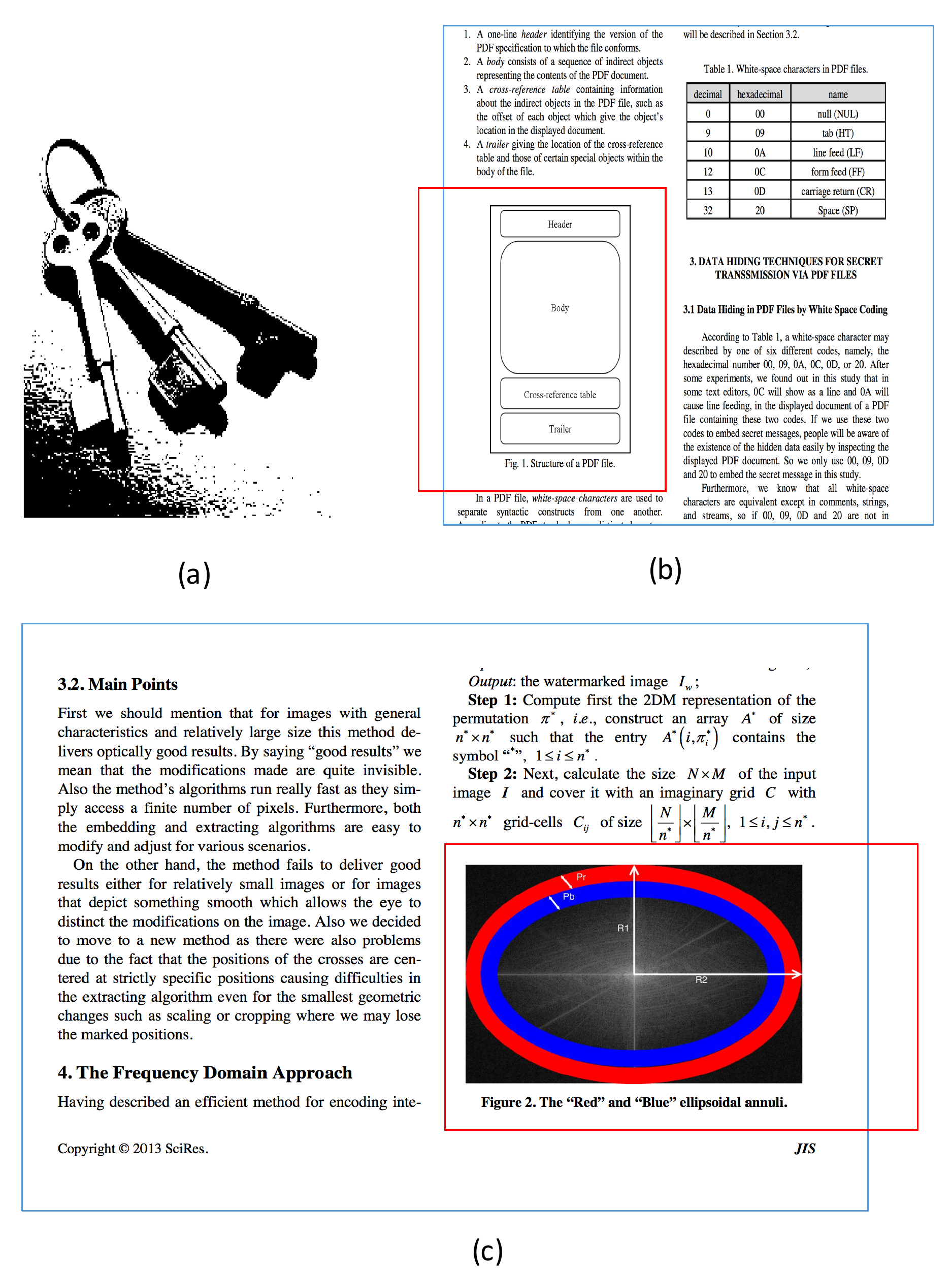}
 \end{center}
 \caption
 { Frequency watermarking of images from PDF document.  a) Binary image that is watermark,
b)  c) two fragments of PDF documents with grayscale and color image.
}
 \label{u3}
 \hyperlink{u3}{W}
\end{figure}

\begin{figure}[!h]
  \begin{center}
    \includegraphics[width=13cm,height=18cm]{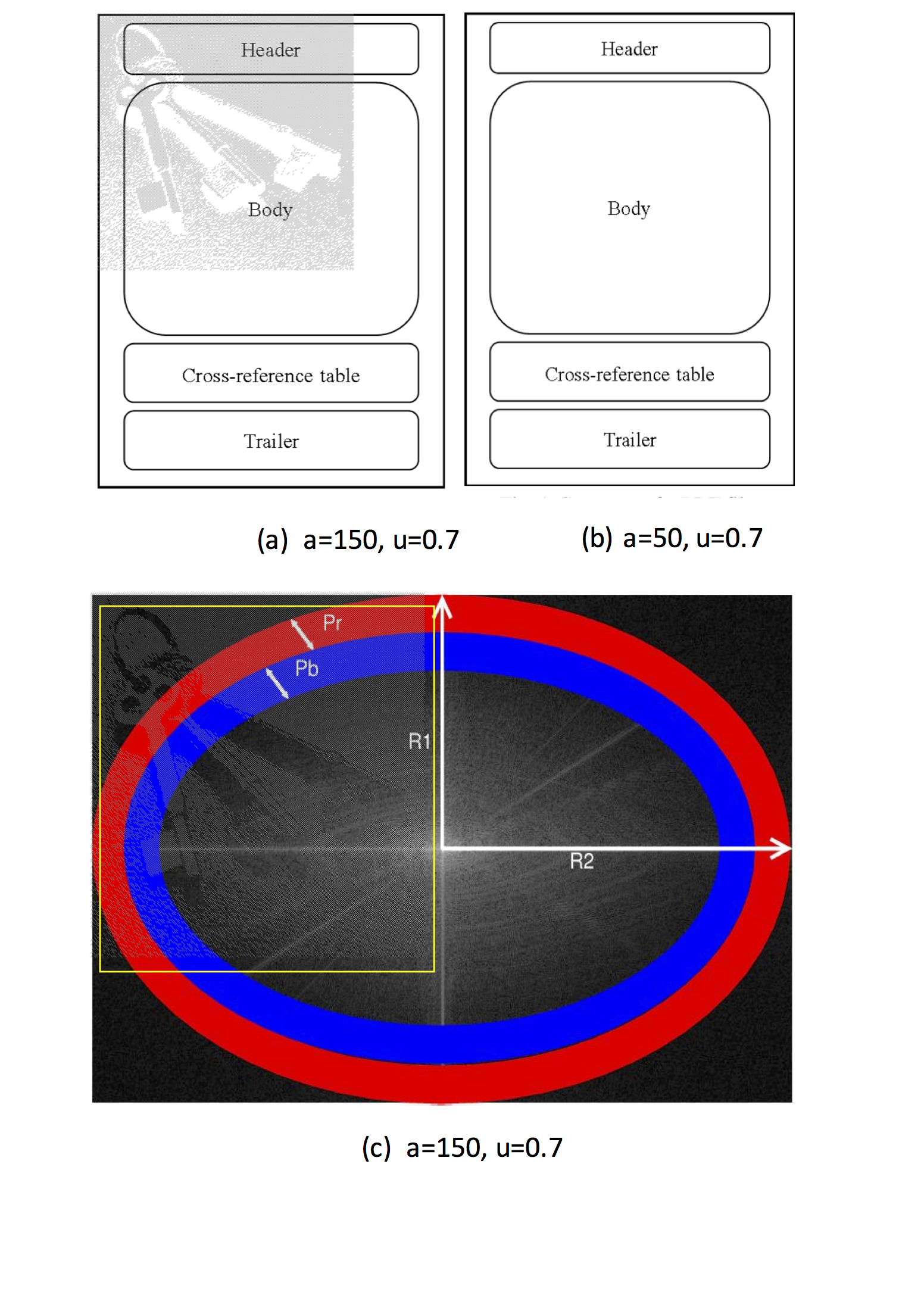}
 \end{center}
 \caption
 { Watermarked Images from PDF pages in Figure (\ref{u3}).  a), b), c) Visible and invisible watermarks.    
}
 \label{u4}
 \hyperlink{u4}{W}
\end{figure}

\begin{figure}[!h]
  \begin{center}
    \includegraphics[width=13cm,height=18cm]{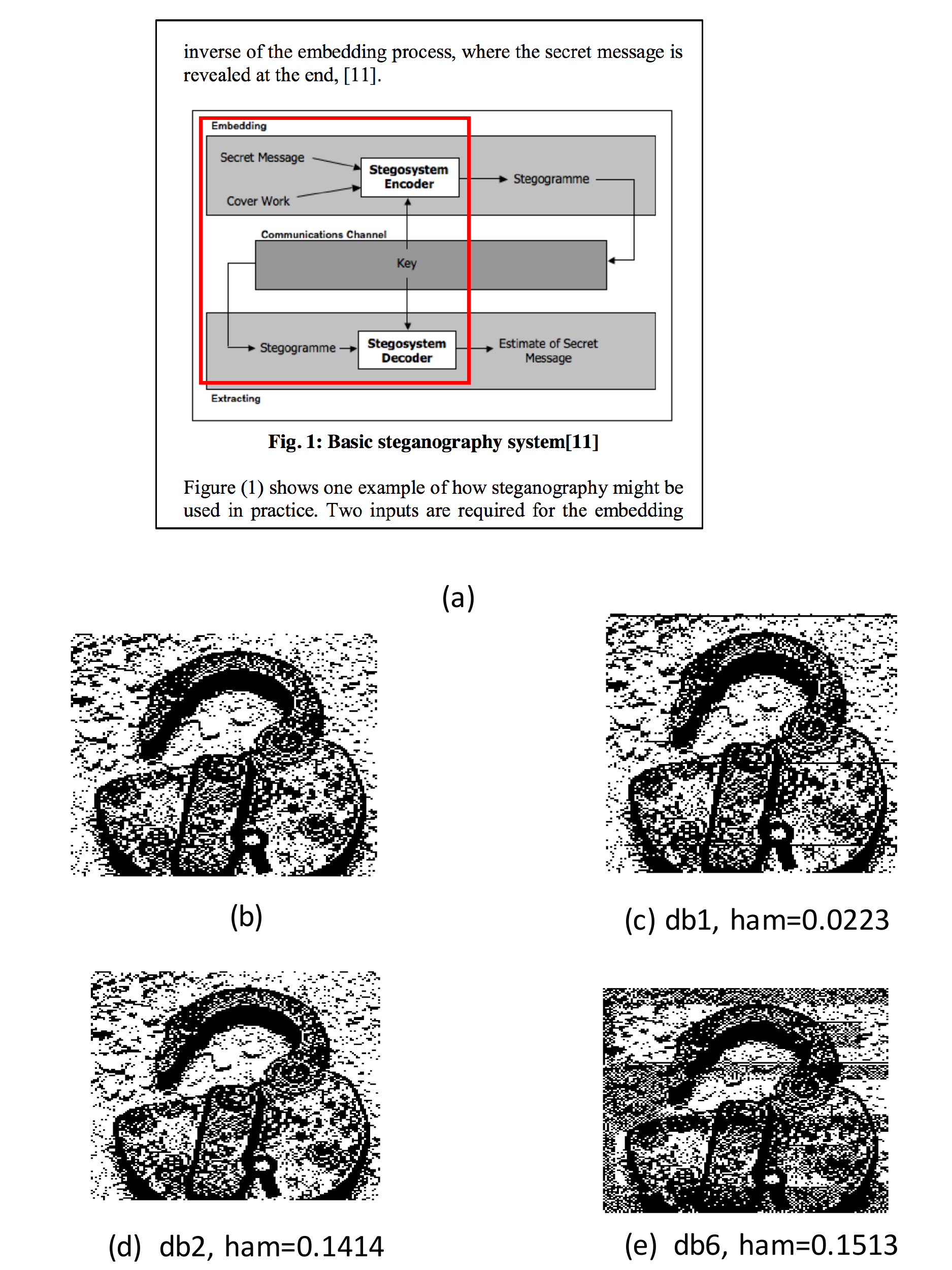}
 \end{center}
 \caption
 { Watermarking by Daubeshies wavelets. a) A fragment of a PDF page with image protected using  invisible watermark, $a=20$, $u=0.5$, \ \ b) watermark,  c), d), e)  extracted watermarks, embedded with the wavelet $db1$, $db2$, $db6$. 
 }
 \label{u5}
 \hyperlink{u5}{W}
\end{figure}

\begin{figure}[!h]
  \begin{center}
    \includegraphics[width=13cm,height=18cm]{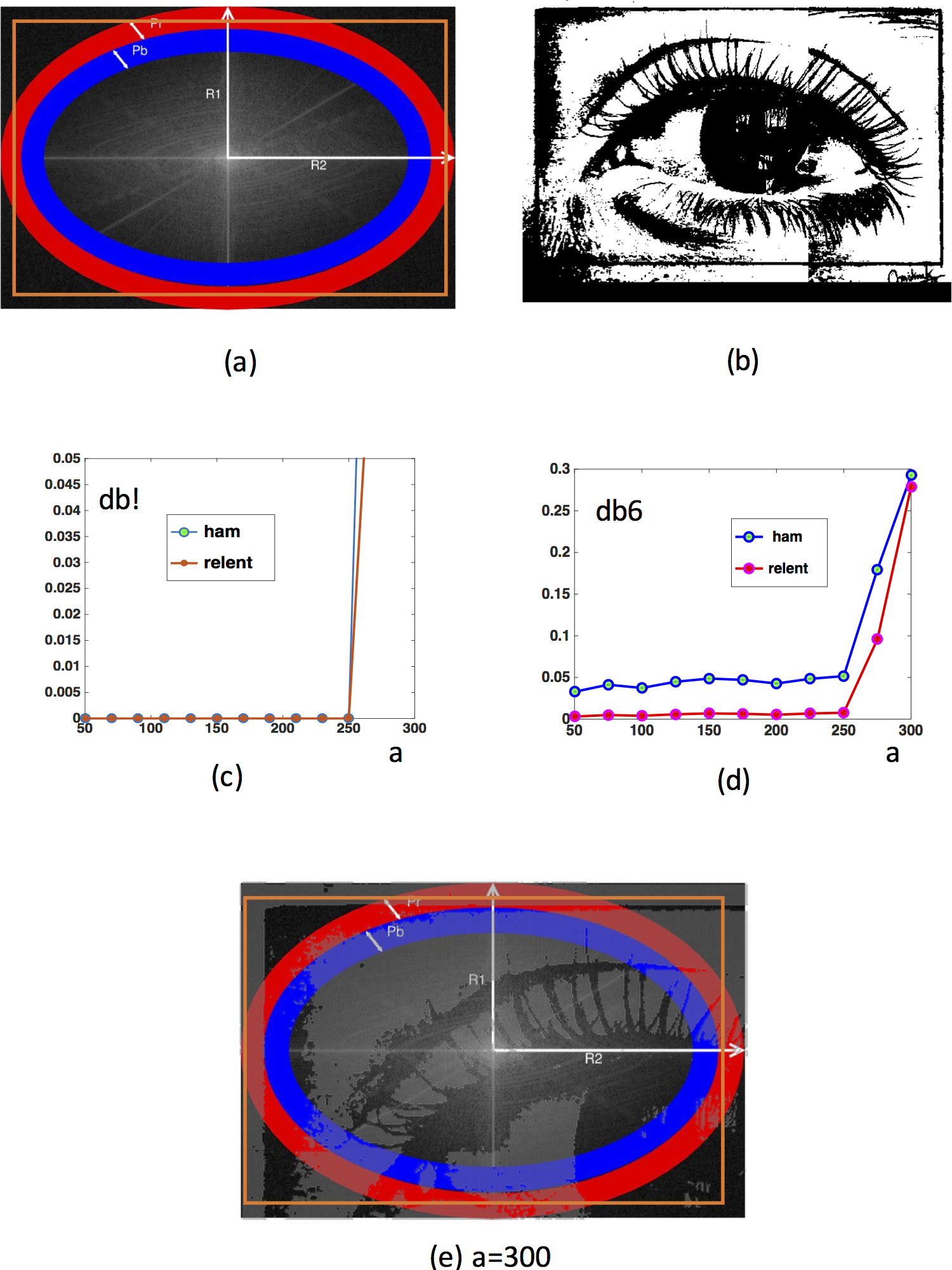}
 \end{center}
 \caption
 { Distortion measures. Hamming distance, \texttt{ham}, and relative entropy, \texttt{relent}, vs brightness of watermarks. a) Cover image, b) binary watermark,  c)  and  d) distortion measures for wavelet db1 and db6, e)  watermarking with hight brightness a=300. 
 }
 \label{u6}
 \hyperlink{u6}{W}
\end{figure}

For frequency domain watermarking  a one level DWT with orthogonal wavelets was used. 
\\
\\
\textbf{Algorithm.} 
\hypertarget{u2}
The frequency embedding scheme is presented in Figure (\ref{u2}). 
The algorithm has the next steps.
\begin{itemize}
\item Transform  cover image $C$ into four blocks of DCT coefficients $cA$, $cH$, $cV$  and $cD$ known as approximal, horizontal, vertical and diagonal details or $LL$, $LH$, $HL$ and $HH$ frequency bands.
\item Choose   a block $Z$ from $cA$, $cH$, $cV$  and $cD$ and replace a part of its coefficients with $M$.
\item  Create the watermarked PNG image $S$ using inverse DWT.
\end{itemize}
To detection $M$ we need to inverse steps.
In Figure (\ref{u2}) (a) any transformation of $S$ is denoted as $T: S\to S'$. This operator considers storing image in PNG file, sending it and converting. Generally $S\neq S'$ because of the introduced errors.  
\\
The embedding algorithm, Figure (\ref{u2}) (b),  can replace a part, $u$, of the block $Z$ with binary image $M$ 
\begin{eqnarray*}
uZ\to aM
\end{eqnarray*}
where $a$ is a scaling parameter, describing brightness of the watermark. Let us assume, that $u=0.9$, $Z=cD$, and $a=2$, so $90\%$ coefficients of the block $cD$ will be replaced with $M$ whose brightness  is increased twice. The brightness of watermark plays an  important role. The introduced changes depend on $a$ and may result in  a visible or an invisible watermark. To get an invisible watermark we need a low value  $a$ that can be established experimentally. 
\\
\
\textbf{Experiment.}  A scheme is focused on invisible watermarks presented by a binary image. The main parameters are  a cover image  and a watermark, the type of wavelet and scaling parameters,   
$\{C,M, w, a,u\}$.  We used the orthogonal wavelets of Daubeshies family $db$ and $sym$ that have good features as for smoothness. For embedding   the $Z=cD$ block was chosen.
\begin{itemize}
\item \textbf{Brightness of watermark.} 
The Figure (\ref{u3}) presents a binary watermark (a)  
\hypertarget{u3}
and two fragments of  PDF document that include a grayscale and a color  images (b) and (d). 
\footnote{These fragments are from paper by
Y-C Lai, W-H Tsai, Covert communication via PDF files by new data hiding techniques, NSC project No, 97-2631-H-009-001and from \cite{PermutwmFreq}} The watermarked images are shown in Figure (\ref{u4}). They 
were obtained by
 the wavelet db1 and $u=0.7$. 
\hypertarget{u4}
The images were extracted from pdf-documents after embedding with $a=50$ and $a=150$.  In the case of  $a=150$ the watermark is visible. We can choose a lower value  $a$ to get invisible watermark.    
\\
\item \textbf{Type of wavelet.}  
Figure (\ref{u5}) illustrates watermarking by various   
Daubeshies wavelets. 
\hypertarget{u5}
A fragment of the PDF document presented in Figure(\ref{u5}) (a) 
\footnote{A page from paper by
A. A. Ali, Al-H S Saad, New text steganography technique by using ma[ed-case-font, International Journal of Computer Application, v, 63, no, 3, 2013.} 
is protected by an invisible watermark shown in Figure (\ref{u5})  (b). For this case $a=20$, $u=0.5$.The extracted watermarks that were embedded with various wavelets and Hamming distance 
$ham(M,M')$  are presented in Figures  (c ) - (e). We considered $db1$, $db2$, $db6$, $db28$, $db41$ and found the Hamming distance 
to be $ham$ =  0.0223, 0.1414, 0.1513, 0.1102, 0.1016.
It tells us that there were errors and  up to 15\% of pixels were restored   incorrectly.  Nevertheless  extracted watermarks look good. The same  result is true for  wavelets from $sym$ family.
\item \textbf{Detection errors.} Degradation of watermarks is caused by   nonreversible transformations and may depend on all the  parameters $\{C,M, w, a,u\}$. 
We calculated a set of  distortion measures to study degradation from  parameter $a$ that is the brightness of embedded watermarks.  Hamming distance, $ham$, and relative entropy, $relent$, are presented in 
Figure (\ref{u6}) for wavelets $db1$ and $db6$.  
The measures tell  that the larger $a$ is the greater  degradation.  
\hypertarget{u6}
For wavelet $db1$ errors given by $ham$  equal about 
 0.003\%  and cause for $db6$ about  3\% . 
\\
\\
It was found that the  error is 0  if watermarked images weren't stored in PNG format before detection.  

\end{itemize}

\section{Spatial embedding}

\begin{figure}[!h]
  \begin{center}
    \includegraphics[width=13cm,height=18cm]{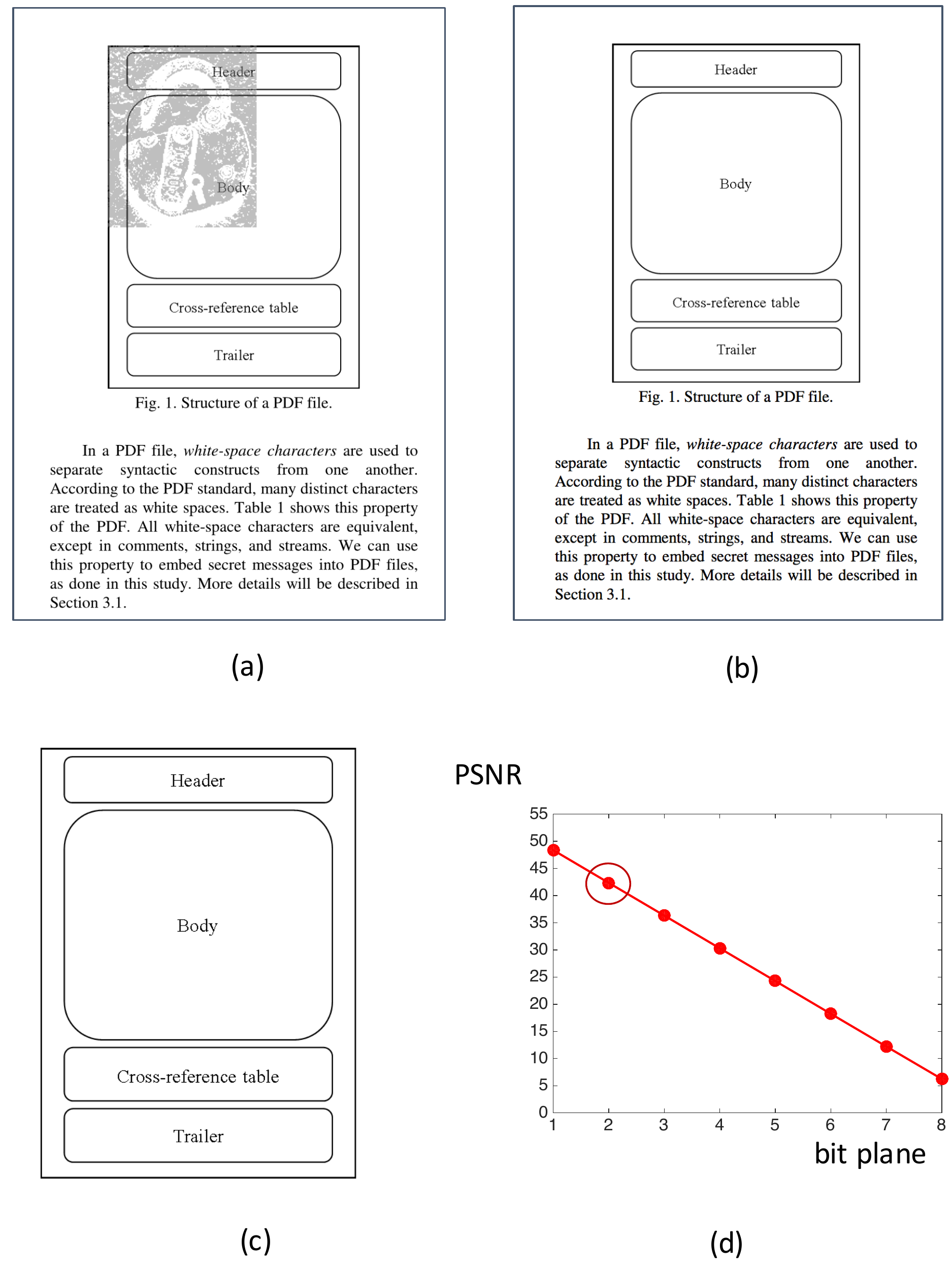}
 \end{center}
 \caption
 { Visible and removal watermark. a) Watermarked image in PDF document, 
$V=7$, $U=2$,  
 b) the image placed in  PDF after removing the watermark,  c)  fragment of retrieved image,   d) PSNR of cover image witout a bit plane $U=1,2,\dots 8$,  
 }
 \label{u7}
 \hyperlink{u7}{W}
\end{figure}

\begin{figure}[!h]
  \begin{center}
    \includegraphics[width=13cm,height=18cm]{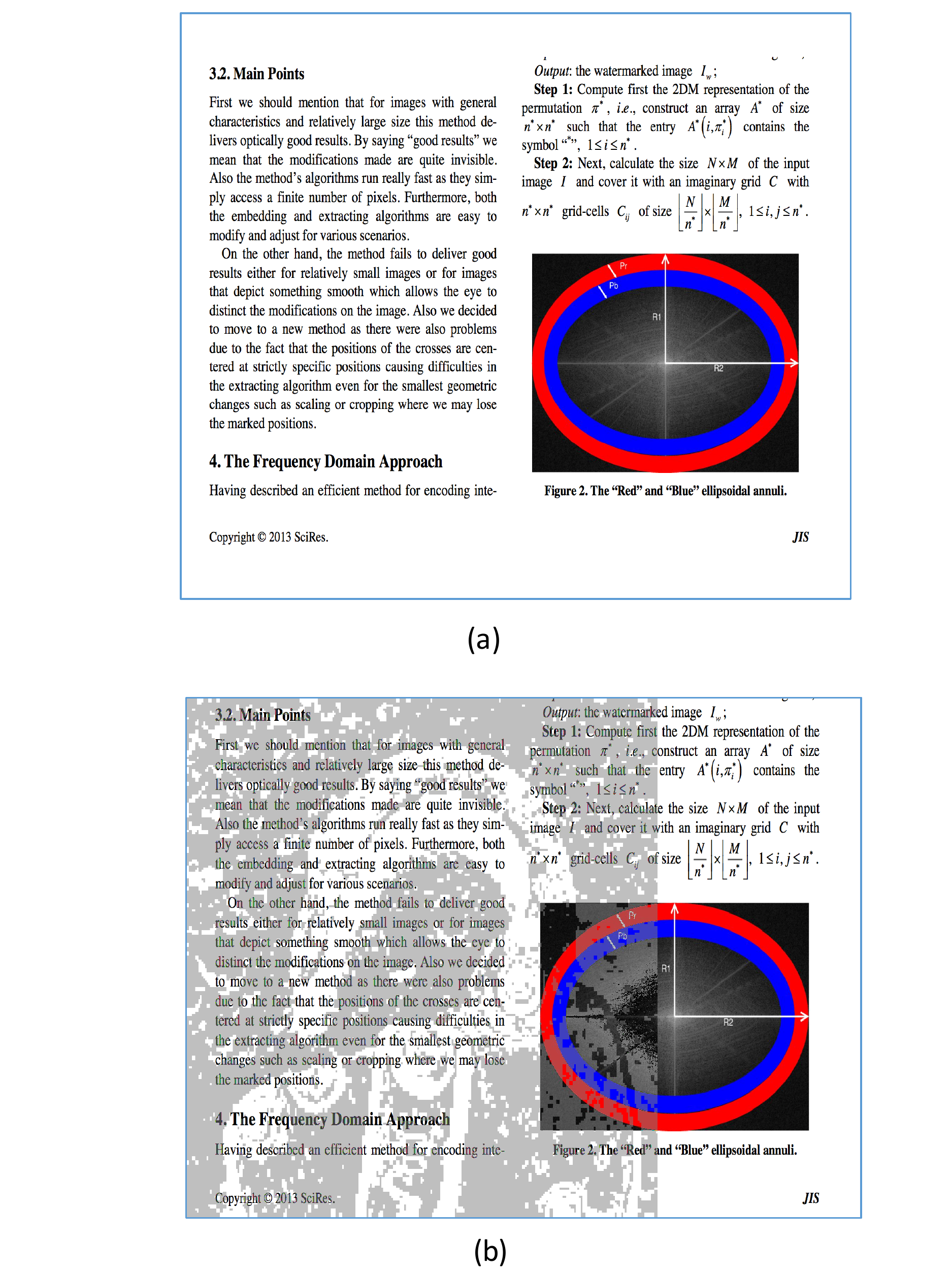}
 \end{center}
 \caption
 { Watermarking of PDF text. a) A fragment of PDF text, 
 b)  the watermarked fragment. 
 }
 \label{u8}
 \hyperlink{u8}{W}
\end{figure}

For spatial embedding the cover image bit planes can be used to get visible and invisible watermarking.
\\
\\
\textbf{Algorithm.} The main idea of  our algorithm is to use a cover grayscale image that has two identical bit planes \cite{Gray}.  This solution allows us  to apply  blind detection of invisible watermarks and  allows us to retrieve the original image after removing the visible watermark. 
\\
Any 8-bit grayscale image $C$ can be represented by its bit planes $B_{V}$, $V=1,\dots,8$. 
Each plane has its weight $2^{V-1}$ and has its semantic information. Let the  least significant  bit plane $B_{U}$, say $U=1,2$, be replaced with  $B_{V}$, where $V>U$. Then the obtained image $C_{2}$ has two identical planes $B_{V}$
\begin{eqnarray*}
C_{2}: \  \mathtt{bitget}(C_{2},V)=\mathtt{bitget}(C_{2},U)=B_{V},
\end{eqnarray*}
where the function \texttt{ bitget} calculates  a given bit plane of a grayscale image. 
\\
\\
The embedding algorithm has two steps:
\begin{itemize}
\item create a cover image  $C_{2}$ with two planes $B_{V}$,
\item  add  a binary watermark $M$ to bit plane $B_{V}$ by modulo 2 
\begin{eqnarray*}
C_{2}\to S=C_{2}-B_{V}2^{V-1}+(B_{V}\oplus M)2^{V-1}.
\end{eqnarray*}
\end{itemize}
Invisible watermark can be achieved, if the embedded plane $B_{V}$ is not a significant plane,  e.g.  $V=2,3$. The detection is blind, it needs the stego image only
\begin{eqnarray*}
S\to M=\mathtt{bitget}(S,V)\oplus \mathtt{bitget(S,U)}.
\end{eqnarray*}
As for a visible watermark we need to choose a significant bit plane $B_{V}$, e.g. $V=7,8$. To remove the watermark the embedded plane is replaced with its copy
\begin{eqnarray}
\label{CRR}
S\to CR=S-\mathtt{bitget}(S,V)2^{V-1}-\mathtt{bitget}(S,U)2^{U-1}+\mathtt{bitget}(S,U)2^{V-1}.
\end{eqnarray}
Clear, that the difference between the cover image $C$ and $CR$ 
is the least significant plane $B_{U}$. This is the main source  of errors.  It may be extremely small  and visually we may  have $C\approx CR$. 
 \\
 \\
\textbf{ Experiment.}  
\begin{itemize}
\item \textbf{Invisible watermark}.The considered spatial algorithms keep the image in the 8 bit integer encoding. There is no loss of information when digital image is stored in the PNG format. In  case of invisible watermarks we can get  no degradation of the extracted data by keeping all the transformations  reversible. 
\item \textbf{Visible and removal watermark.}
\hypertarget{u7}
Figure (\ref{u7}) shows the watermarked fragment of a PDF document. The bit plane $V=7$ was used. This plane was storied earlier as the least significant plane $U=1$.  
Figure (\ref{u7})  (b) and ( c)  present the retrieved images $CR$ placed in his 
PDF documents and its digital versions. Both retrieved images  look nice and they are visually undistinguished from their originals. 
To find a difference between $C$ and $CR$
we consider PSNR. In accordance with 
(\ref{CRR}) the difference is defined as $B_{U}$.
Figure (\ref{u7}) (d) presents PSNR calculated between cover image $C$ and $C_{U}$  that  is the cover  image without plane $U=1,2,\dots$,  In our case $U=2$ and  PSNR=42.3417 db. This corresponds to PSNR between $C$ and $CR$ that was found from experiment.
Note, the large  PSNR is in agreement with visual quality.  As it follows from 
the Figure (d), PSNR of about 50 db is achieved if the cover bit plane is copied into the least significant plane $U=1$.  
\item 
\textbf{Watermarking of text.}
Our convertor has an option to recognize a PDF page with text as an image. Then the PDF text is converted into PNG format and can be watermarked with the help of the considered technique. 
As a result the PDF  text is  protected as the  Figure (\ref{u8}) shows.  
\end{itemize}

\section{Conclusions}
For frequency domain embedding we analused DWT and found that the main reason of  errors 
is averaging when the digital image is storied in a graphical format. We used PNG format requiring 8-bit integer encoding.
\\  
The introduced spatial domain embedding  turned  out to be free of such errors. This technique is based on bit planes of cover images and allows to embed  invisible and visible watermarks. We introduce an algorithm that can double  the bit planes of  gray scale cover image. As a result a visible watermark can be removed and the cover image can be retrieved with high visual quality.   


\newpage


\end{document}